# Electro-Hydrodynamic Shooting Phenomenon of Liquid Metal Stream

Wen-Qiang Fang [a)], Zhi-Zhu He [b)*] and Jing Liu [b,c)*]

[a)] Department of Engineering Mechanics, School of Aerospace Engineering, Tsinghua University, Beijing 100084, China

[b)] Key Laboratory of Cryogenics, Technical Institute of Physics and Chemistry, Chinese Academy of Sciences, Beijing 100190, China

[c)] Department of Biomedical Engineering, School of Medicine, Tsinghua University, Beijing 100084, China

**\*Address for correspondence:**
Dr. Zhi-Zhu
Technical Institute of Physics and Chemistry,
Chinese Academy of Sciences,
Beijing 100190, P. R. China
E-mail address: zzhe@mail.ipc.ac.cn
Tel. +86-10-82543766
Fax: +86-10-82543767

Dr. Jing Liu
Technical Institute of Physics and Chemistry,
Chinese Academy of Sciences,
Beijing 100190, P. R. China
E-mail address: jliu@mail.ipc.ac.cn
Tel. +86-10-82543765
Fax: +86-10-82543767

**Abstract:**

We reported an electro-hydrodynamic shooting phenomenon of liquid metal stream. A small voltage direct current electric field would induce ejection of liquid metal inside capillary tube and then shooting into sodium hydroxide solution to form discrete droplets. The shooting velocity has positive relationship with the applied voltage while the droplet size is dominated by the aperture diameter of the capillary nozzle. Further, the motion of the liquid metal droplets can be flexibly manipulated by the electrodes. This effect suggests an easy going way to generate metal droplets in large quantity, which is important from both fundamental and practical aspects.

**1. Introduction**

In recent years, the room temperature liquid metal has attracted much attention because of their versatile applicability in energy management [1, 2], chip cooling [3] and printed electronics [4]. A lot of unique characters involved are thus increasingly investigated [5-9]. Among the many issues ever tackled, the production of liquid metal droplets or particles with controlled size has been identified to be very useful in a wide variety of important areas. Typical examples can be found in MEMS [10], liquid marble preparation [11] or microfluidic pump [12, 13] etc. So far, several important approaches have been developed to produce the liquid metal droplet in micro-channel [14, 15]. In those works, the droplets take shapes by flow focusing, and the key factors for controlling the fabrication include fluid velocity, viscosity and surfactant properties. Due to pre-requisite in the manufacture of the micro-fluidic channels, such method is still somewhat expensive and technically complex. For a smaller size, the liquid metal microspheres can even be prepared to the nanoscale based on ligand mediated self-assembly method [16]. In a latest work, a straight forward way was found for large-scale fabrication of liquid metal micro-droplets and particles [17]. The mechanism there lies in the Plateau–Rayleigh instability [18], where a liquid jet would break up into smaller packets because of the high surface tension of the liquid metal inside the matching solution. As is noted, the mechanical manipulation mechanism there is still not convenient enough for a continuous fabrication of the metal droplets.

Through continuous efforts, we found in the present work that the mechanical ejection can in fact be replaced by an electro-hydrodynamic effect. It is based on this fundamental discovery that, we reported an alternative way of generating liquid metal droplets through the electrically controlling mechanism. The disclosed process and device are rather flexible and easy going. Given automatic control, this method would significantly improve the fabrication efficiency of the liquid metal droplets in the coming time.

## 2. Materials and Methods

To carry out the experiments, we have set up the test platform as shown in Fig. 1(A) with working mechanisms illustrated in Fig. 1(B) and (C), respectively. Here, the capillary tube serves as the channel connecting the liquid metal and the sodium hydroxide (NaOH) solution container, where the cathode and anode are arranged as depicted in the figure. Two diameters of the capillary tube as 1mm and 0.7mm were comparatively studied. The cathode and anode made of stainless steel are both linked with the direct-current (DC) voltage controller. Regarding the test liquid metal, it was chosen as galinstan (made of 67%Ga, 20.5%In, and 12.5%Sn by volume), which has a broad temperature range of liquid phase with a melting point at 10.35 ℃ [19].

We firstly adjust the height of liquid metal level of the container so that the liquid metal can be infused into the capillary tube which however cannot flow out of the nozzle due to its pretty large surface tension. The practical distance of the two electrodes from the capillary nozzle to the anode is about 82mm because of the conductive characteristics of the liquid metal. The voltage controller is turned on to apply the DC electric field on the electrolyte solution. Then, an unconventional phenomenon was discovered that the liquid metal would automatically eject from the capillary nozzle, then shoot into the electrolyte solution, and form droplets until finally move to the anode. The whole process is recorded by a high speed camera (IDT, NR4.S3). The velocity of droplets can thus be calculated from the videos through image processing. Through altering the voltage, concentration of solution and the aperture size of capillary tube, we could investigate the effects of various typical factors on the droplets generation behavior.

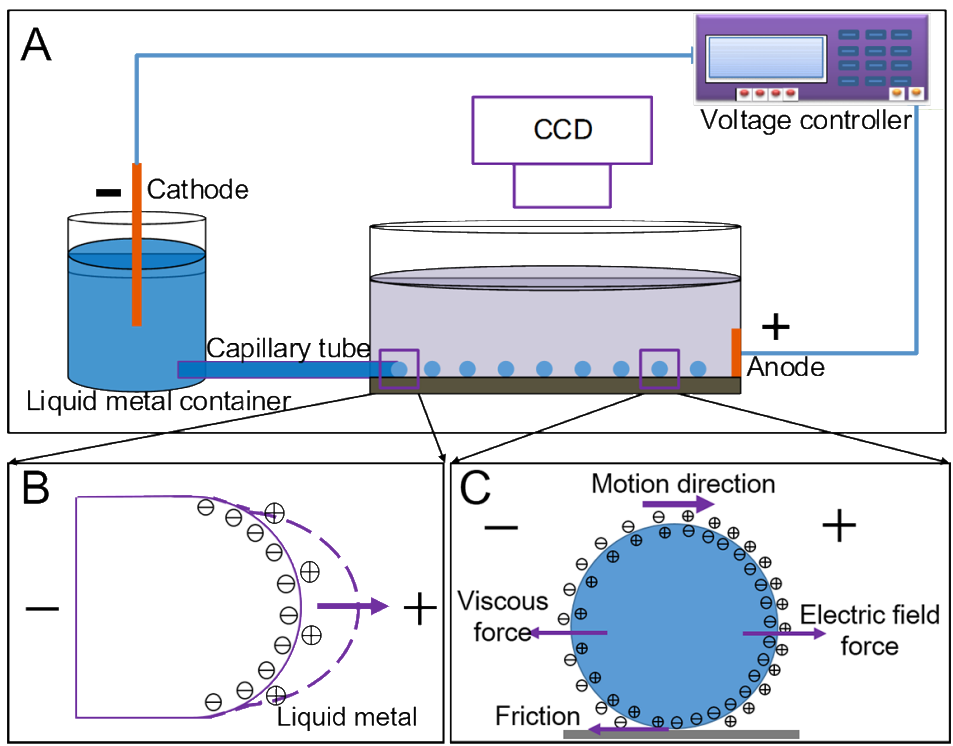

FIG. 1. (A) The schematic diagram of the experimental setup. Both of the liquid metal injection (B) and droplet motion (C) are driven by electro-hydrodynamic force.

## 3. Results

When electric field is applied to the electrolyte solution, the force balance between pressure and surface tension on the interface of liquid metal and NaOH solution at the capillary nozzle is broken immediately. The traction force induced by the external electric field would then promote the liquid metal to eject from the capillary nozzle and shoot into the electrolyte solution. Due to large surface tension of the liquid metal, the stream then splits to form a large amount of the droplets continually. Fig. 2 shows the snapshots of typical ejections in NaOH solution of 0.25mol/L under voltages from 2.5V to 20V. The intensity of the electric field can be considered as linear dependence on the applied voltage. For the too much low voltages (below 2.5V), we did not observe the liquid metal droplet generation due to its high surface tension. When raising the voltage strength, the injection velocity of the liquid metal increases evidently. Overall, the ejection direction of the liquid metal is along the central axis of the nozzle for the voltages below 5V (Fig. 2(A-B)). However, it is interesting to note that such ejection direction becomes unstable, which is affected by the high voltage (Fig. 2(C-E)). Turbidity around the cathode was seen when the voltage strength increases to about 5V. These dark-grey matters are composed mainly of compounds containing In and Sn ions due to the electrochemical reaction at the interface between liquid metal and NaOH solution. From Fig. 2, it is also found that the droplets are attracted to the anode under external electric field.

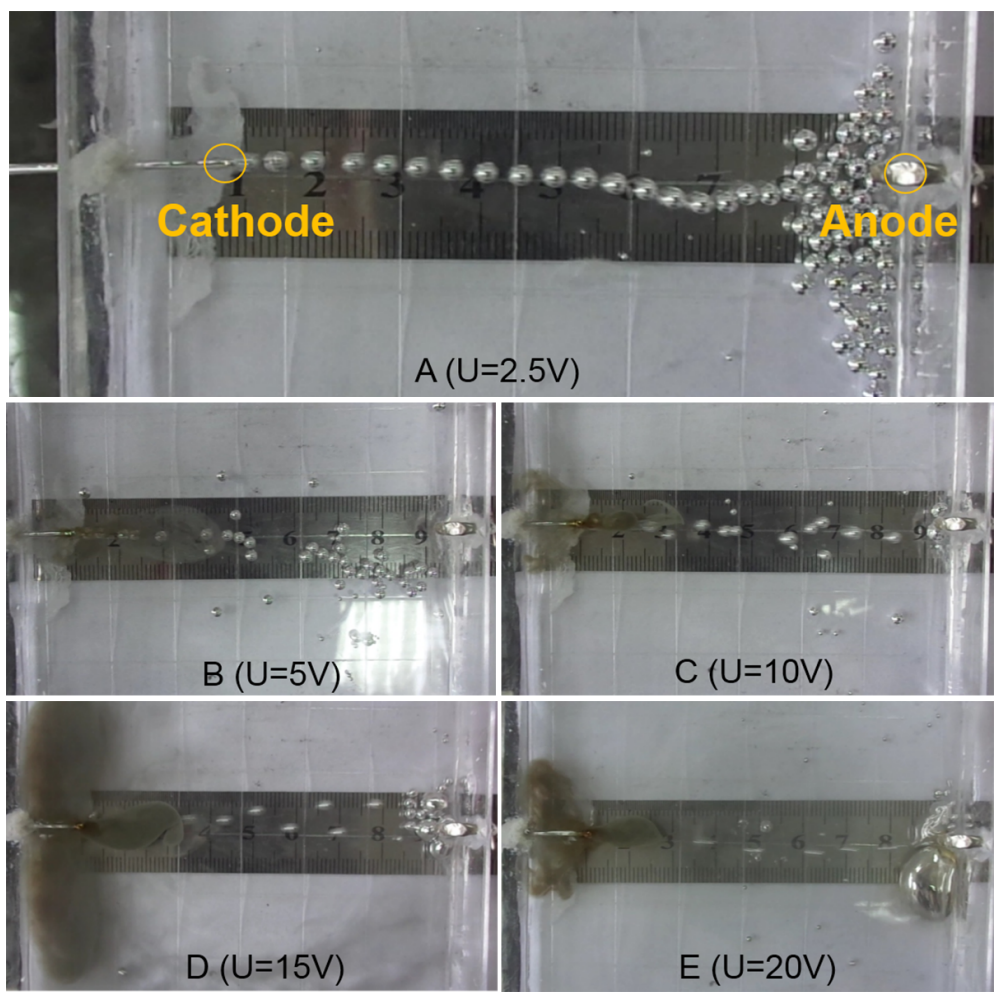

FIG. 2. The snapshots of liquid metal shooting in NaOH solution of 0.25mol/L under different voltages: (A) U=2.5V; (B) U=5V; (C) U=10V; (D) U=15V; (E) U=20V.

The liquid metal droplets' generation behavior is mainly dominated by the size of the capillary nozzle, the voltage and the concentration of the NaOH solution. Fig. 3 depicts the relation between the applied voltage and liquid metal droplet velocity for different aperture sizes of capillary nozzle and positions in 0.125mol/L NaOH solution, where the position is denoted by the distance from the capillary nozzle. From the measurements, one can conclude that the ejection speed of the droplets goes up rapidly with the increment of the voltage. The velocity of the droplet decreases due to viscous resistance effect applied on it during traveling along the solution. Fig. 3 also indicates that the dependence of the droplet velocity on the voltage is less affected by the aperture size of the capillary nozzle considered here.

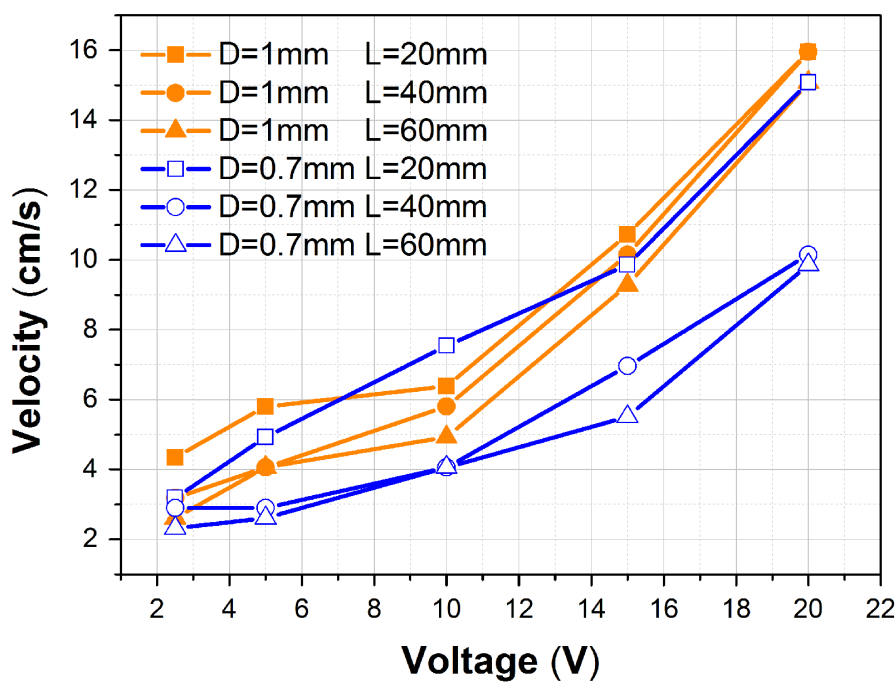

FIG. 3. The relationship between applied voltage and liquid metal droplet velocity for different aperture sizes of capillary nozzle (D) and positions (L denotes the distance from the capillary nozzle) in 0.125mol/L NaOH solution.

Fig. 4 presents the relationship between velocity and concentration of NaOH solution under applied voltage from 5V to 20V at the position (2cm away from the capillary nozzle). The concentration of NaOH solution has no significant effect on the velocity of the droplets. In fact, increasing the concentration of NaOH would lead to the decrease of the electric permittivity of the electrolyte solution [20], and weaken the electro-hydrodynamic driving force. However, this effect is not evident in the present experiments. Besides, the relationship between voltage and velocity does not generate prominent difference with different capillary nozzles. According to the experiments, the size of the liquid metal droplets is mainly determined by the capillary nozzle diameter. When using capillary of diameter 1mm, the average size of liquid metal droplet is about 2mm. And for the case of diameter 0.7mm, the average size of droplet is about 1.6mm. Further, we also observed that increasing the voltages can slightly lead to smaller droplets. In addition, the inner surface roughness of the nozzle also affects the droplet size.

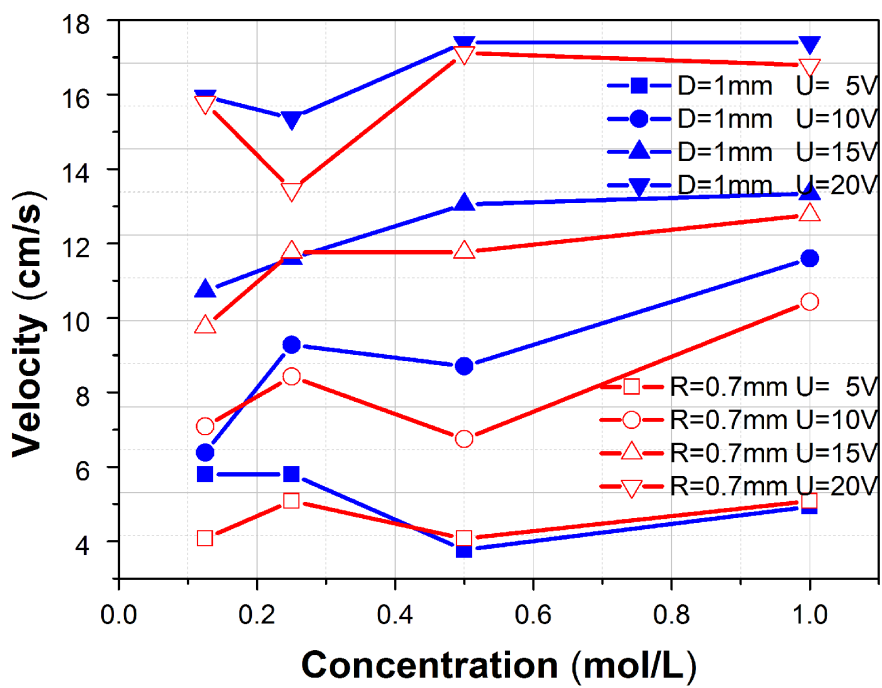

FIG. 4. The relationship between velocity and concentration of NaOH solution under different applied voltages at the position 2cm away from the capillary nozzle.

**4. Discussion**

Theoretically speaking, the present finding regarding the metal droplet generation and manipulation can be attributed to the fundamental electro-hydrodynamic mechanism of the interaction between liquid metal and electrolyte solution. Overall, the whole process can be divided into three phases: liquid metal ejection induced by electric field, the liquid metal stream breaking into droplets and the droplets motion in the base solution.

The first phase provides the initial momentum of the ejection in analogy to external mechanical force [17]. Without losing any generality, the Yang-Laplace equation $\Delta p = 2\gamma/R$ can be used to characterize the force balance on the interface of liquid metal at initial stage, where $\Delta p$ is the pressure difference between liquid metal and base solution and determined by the both liquids level, $\gamma$ is the surface tension of liquid metal (0.718N/m for galinstan [4]), and $R$ the radius of liquid metal sphere. For $R$=1mm, the liquid metal sphere can sustain a large pressure difference 1436 Pa. When the electric field is applied, an electrical double layer (EDL) is formed at the interface of the liquid metal (Fig.1(B)). The induced electric force can be denoted by $\varepsilon E^2$, where $\varepsilon$ is the electric permittivity of NaOH solution and E the electric filed strength, acted on the liquid metal interface along its normal direction. As a result, the equilibrium of surface tension and pressure is broken. Then the interface deforms and tends to move toward the side of base solution. After this acceleration process, the liquid metal ejects out from the nozzle. Obviously, increasing the applied voltage will result in a larger electric force.

In the second phase, a liquid metal stream breaks into droplets due to Plateau–Rayleigh instability. It should be mentioned that, we did not observe here the continuous thin stream travelling phenomenon as found in the mechanical force controlled liquid metal injection [17]. The reason lies in that the electro-hydrodynamic force has much stronger effect on the flow instability, which thus enhances the liquid metal droplet generation and leads to the disorder of the injection direction.

For the third phase, the liquid metal droplet is driven by electro-hydrodynamic force to move along the proposed direction through the electrodes layout, which can be adopted for precise manipulation of the droplets. The basic phenomenon can be understood from Fig.1(C). Immediately after external electric filed is applied, the current then drives positive ($Na^+$) and negative ions ($OH^-$) to move towards the corresponding side of the liquid metal droplet, which induces an equal and opposite surface charge on the conducting surface. Thus the double-layer charge density under the electric field would induce the droplet motion. Besides, droplets carrying negative charges when ejecting out of the capillary at cathode further contribute to the electric field force. Based on the electro-hydrodynamic theory [21], the velocity of the liquid metal droplet can be deduced as:

$$U = \frac{9kD}{40(1+\mu_L/\mu_W)} \frac{\varepsilon D E^2}{\mu_W} \tag{1}$$

where, $k^{-1}$ is the Debye length (about $5.0\times10^{-9}$m), $D$ is the diameter of the liquid metal (about $2^{-3}$m), $\mu_L$ is the viscosity of galinstan (2.4×10$^{-3}$ Pa s at 20 ℃ [19]) and $\mu_W$ for NaOH aqueous solution (about 1.0×10$^{-3}$ Pa s at 20 ℃), $\varepsilon$ is the electric permittivity of NaOH aqueous solution (about 6.75×10$^{-10}$ Fm$^{-1}$ for 0.125mol/L [20]). *E* denotes the electric filed strength chosen as 61 V/m for voltage 5V. Thus the velocity of the liquid metal droplet estimated from Eq. (1) is about 13.3cm/s, which is higher than the experimental results about 3cm/s. The reason for this deviation lies in that the Eq. (1) is derived from the balance between viscous force and electric field force in free space. However the friction from the current substrate impedes the droplet motion. In addition, the electrochemical reaction on the liquid metal droplet surface could induce the surrounding flow disorder, and thus weaken its directional motion. According to Eq. (1), the velocity of the droplet depends linearly on the electric permittivity. For 1mol/L NaOH solution, its electric permittivity is [20]: 5.70×10$^{-10}$ Fm$^{-1}$, which does not have too much difference with that of concentration 0.125mol/L. Thus, droplet velocity depends less on the concentration of the NaOH solution as considered here.

The velocity of metal solid particle induced by the external electric field in electrolyte solution is given by $U = \varepsilon D E^2 / \mu_W$, and estimated as 5.0 um/s according to the above parameters, which is much smaller than that for liquid metal droplet. The reason lies in that the tangential electric field vanishes at free surface of liquid metal droplet. The viscous stress associated with Debye-scale shear within the electrolyte must be balanced by the electric stresses, which leads to amplified velocity scaling about $kD$ compared with the metal solid particle. It is noteworthy that the NaOH solution plays a key role for droplet motion with high velocity. The liquid metal surface tends to come into being $Ga_2O_3$ due to electrochemical reaction under electric field, which decreases the surface tension and liquidity, and weakens the electro-hydrodynamic effect. However, NaOH solution can effectively deoxidize $Ga_2O_3$. For NaCl solution, the velocity of the liquid metal appears smaller than that for NaOH solution.

## 5. Conclusion

In summary, we have discovered a fundamental electro-hydrodynamic phenomenon that low magnitude electric field would easily induce liquid metal ejection from a capillary tube. The subsequent shooting of the metal stream into the solution would generate a large amount of discrete droplets. The carried out experiments disclosed the major factors to dominate the events. Several important conclusions can be drawn as follows. Firstly, the ejection velocity of galinstan droplets has positive correlation with the applied voltage. Secondly, the concentration of NaOH solution has no significant effect on the ejection velocity. Thirdly, the size of the galinstan droplets depends mainly on the aperture diameter of the capillary nozzle. The present finding opens an efficient strategy to flexibly fabricate liquid metal droplets in large amount and with controlled size via a rather rapid, easy and low cost way. It also raised important scientific issues worth of investigation in the coming time.

**Acknowledgments:**

This work is partially supported by the by the Research Funding of the Chinese Academy of Sciences (Grant No. KGZD-EW-T04-4) and NSFC under Grant 81071225.